# Photoluminescence of MAPbI$_3$: a semiconductor science and technology point of view.

Valerio Campanari[1*], Antonio Agresti[1], Sara Pescetelli[1], Aswathi K. Sivan[2], Daniele Catone[3], Patrick O'Keeffe[4], Stefano Turchini[3], Aldo Di Carlo[1,3], and Faustino Martelli[2*]

[1] CHOSE (Centre for Hybrid and Organic Solar Energy), Department of Electronic Engineering, University of Rome Tor Vergata, Via del Politecnico 1, 00133 Rome, Italy

[2] CNR-IMM, Area della Ricerca di Roma Tor Vergata, 100 Via del Fosso del Cavaliere, 00133 Rome, Italy

[3] Istituto di Struttura della Materia – CNR (ISM-CNR), , Division of Ultrafast Processes in Materials (FLASHit), Area della Ricerca di Roma Tor Vergata, Via del Fosso del Cavaliere 100, 00133 Rome, Italy

[4] Istituto di Struttura della Materia – CNR (ISM-CNR), , Division of Ultrafast Processes in Materials (FLASHit), Area della Ricerca di Roma 1, 00015 Monterotondo Scalo, Italy

## ABSTRACT

In this work, we perform steady-state continuous wave (cw) photoluminescence (PL) measurements on a MAPbI$_3$ thin film in the temperature range of 10-160 K, using excitation densities spanning over almost seven orders of magnitude, in particular investigating very low densities, rarely used in the published literature. The temperature range used in this study is below or at the edge of the orthorhombic-tetragonal phase transition in MAPbI$_3$. In particular, we show that even in high quality MAPbI$_3$, capable of providing high photovoltaic efficiency, the defect density is high enough to give rise to an energy level band. Furthermore, we show that the intensity ratio between the two PL components related to the two crystalline phases, is a function of temperature and excitation. At high excitation intensities, we show that amplified spontaneous emission is attainable even in cw conditions. Time-resolved PL is also performed to justify some assignments of the PL features. Finally, our systematic approach, typical for the characterization of semiconductors, suggests that it should also be applied to hybrid halide perovskites and that, under suitable conditions, the PL characteristics of MAPbI3 can be reconciled with those of conventional inorganic semiconductors.

# I. INTRODUCTION

The employment of hybrid halide perovskites (HHPs) as the active material in solar cells has in the last few years undergone a tremendous development [1,2]. This is due to the remarkable properties of these low-cost materials, such as high light absorption over the entire visible range [3], long diffusion length of charge carriers [4], as well as long recombination times that prevent electron-hole recombination for an efficient charge collection [5]. Among the different types of materials, methylammonium lead iodide ($CH_3NH_3PbI_3$, $MAPbI_3$) is one of the most investigated compounds because it is among those with the best properties in terms of charge transport and light absorption and $MAPbI_3$ solar cells reach the best performances.

The most common characterization technique of HHPs is photoluminescence (PL) [6-8] as the emitting properties of a material have always been related to its potential efficiency as a photovoltaic light absorbing material [9]. The number of published works on the PL of HHPs is large and papers that review the results periodically appear in the literature [6-8]. In general, PL measurements on HHPs are performed at room temperature (RT) and over a limited range of excitation conditions [10-15]. If, on the one hand, performing measurements at room temperature is fully justified by the fact that photovoltaic cells operate under ambient conditions, on the other hand the thermal energy at 300 K is often enough to hide the presence of defects in the sample and does not allow a true characterization of the material. For $MAPbI_3$, low-temperature measurements are mainly performed to investigate the phase transition between tetragonal (T) and orthorhombic (O) lattice structures that occurs around 160 K but that does not appear as a sharp feature [7], [16-19]. Also in this case, however, most studies are made over limited excitation-intensity (when given) or temperature ranges [10-12,15,20-26] and often they are carried out using pulsed laser excitations [11,12,20,25,26] which generally induce high densities of photoexcited electron-hole (e-h) pairs such that some recombination channels could remain unobserved due to saturation. This quite unsystematic characterization, combined with the morphological variation and difficult reproducibility of this kind of material, makes the comparison between different works arduous, and hence also between materials produced in different laboratories. For example, although all refs. report the presence of at least two emission peaks below the O-T transition for certain temperature ranges, the energy position and the relative intensities of the peaks can differ between different works. In some cases [10,11] the lower energy peak is dominant for all the temperature range, in others [14,15,21,23] the higher energy peak becomes dominant below a certain temperature.

In this work, we study the PL properties of MAPbI$_3$ samples capable of delivering a photovoltaic conversion efficiency above 20%, when incorporated into a device of mesoscopic n-i-p architecture [27], from the point of view of the well-established methodology of the semiconductor science and technology. With this aim in mind, we measured the PL at low temperatures using a continuous-wave (cw) light source (in the end, the sun is a cw source) with incident excitation densities spanning over almost seven orders of magnitude: starting from the lowest value capable of giving a sufficiently good signal-to-noise ratio (1.3x10$^{-4}$ W/cm$^2$), up to very high excitations (~4x10$^2$ W/cm$^2$) that are well above the common application limits of the material. Indeed, the integrated sun intensity at AM 1.5 is 1 x 10$^{-1}$ W/cm$^2$, which is in the middle-low part of this range. Moreover, perovskites are used also for solar cells working in low light conditions (down to 30 μW/cm$^2$).

The temperature range of our study is 10-160 K at the edge or below the critical temperature for the O-T phase transition. For higher temperatures, the results are less sensitive to the excitation intensity and the comparison of our findings with the enormous amount of data in the literature recorded at RT is not within the scope of the present work. We paid particular attention to low power densities that are rarely used for probing MAPbI$_3$ and, more generally

HHPs, because they allow the observation of low-density electronic states such as those related to crystal defects and impurities or, in the case of MAPbI$_3$, the electronic states due tetragonal inclusions in an orthorhombic lattice [19]. We will show that the presence of contrasting results reported in literature on the O-T phase transition can be explained by the use of different experimental conditions, and we will also point out the effect of defect states on the PL spectra at energies below the band-gap. Time-resolved PL measurements were also performed to provide additional support to our conclusions.

Beyond the characterization of our samples, this study will also stimulate the perovskite community to adopt a common way to characterize HHP samples, in order to provide a common and shared methodology, as it has been suggested for perovskite solar cells stability [28].

## II. EXPERIMENTAL

MAPbI$_3$ samples are prepared as detailed in the following. After a deep cleaning sequence, consisting of a triple step of ultra-sonic bath with cleaning liquid dissolved in deionized water, acetone and 2-propanol for 10 min each, simple glass substrates are coated with a MAPbI$_3$ perovskite absorber deposited by solvent engineering methods. Briefly, 717.76 mg mL$^{-1}$ of PbI$_2$ and 247.56 mg mL$^{-1}$ of CH$_3$NH$_3$I were dissolved in DMF/DMSO (8:1, v/v) by stirring for 24 h at room temperature to obtain the perovskite based solution. 70 μL of the perovskite solution was spin coated on the glass with a two-step spin-coating process, first at 1000 rpm for 10 s and then at 5000 rpm for 45 s. Just 34 s before the end of the second spin-coating step, 0.7 mL of diethyl ether was dropped onto the substrates. Subsequently, the perovskite layer was treated with a double-step annealing process, performed at 50 °C for 2 min in the dark and then at 100 °C for 10 min. The resulting sample have thickness of about 450/500 nm. Photovoltaic cells based on this type of material show efficiency of 20.12% [27]. The PL measurements were performed by exciting the samples with a 405 nm CW laser diode focused to a diameter of approximately 100 μm. The excitation density has been varied from $1.3 \times 10^{-4}$ to $\sim 4 \times 10^2$ W/cm$^2$. The sample was held at temperatures in the range 10-160 K in a close-loop He cryostat. The luminescence has been detected with a Peltier-cooled CCD camera after having been dispersed in a 30 cm long monochromator with a 1200 grooves/mm grating. The time-resolved PL measurements were performed at 77 K with a time-correlated single photon counting (TCSPC) system with a photomultiplier as the detector with an instrumental response function (IRF) of 0.5 ns. The measurements were performed using a pulsed laser source (duration 40 fs) at 520 nm with 1 KHz repetition rate for three different fluences (1.13 μJ/cm$^2$, 5.7 μJ/cm$^2$ and 57 μJ/cm$^2$).

## III.  Results

Before going into the details of our results, some general considerations based on the cited literature can be made. At room temperature, the MAPbI$_3$ perovskite presents one PL peak due to its tetragonal phase bandgap. Lowering the temperature below ~150 K, a new, higher energy peak emerges, attributed to recombination in the orthorhombic phase. Both peaks continue to be visible in the PL spectra of MAPbI$_3$ also at lower temperatures, well below the structural phase transition (160 K). From now on, for simplicity, we will refer to those peaks as LE (low energy, tetragonal phase) and HE (high energy, orthorhombic phase).

*Power dependent photoluminescence with cw excitation.*

Fig. 1(a) presents the PL spectra as a function of the excitation power density ("$I_{exc}$") recorded at 10 K, from 0.13 mW/cm$^2$ up to 380 W/cm$^2$ on the same spot of the sample. The measurements were repeated for another spot (Fig. 1(b)) to compare the results. As already described in the introduction, a low $I_{exc}$ allows the investigation of the material in conditions of sufficiently low photoexcited carrier densities in which well-defined and standard measurements are possible, also avoiding saturation effects.

Moreover, it is well known that halide perovskite thin films very often show macroscopic inhomogeneity. Therefore it is important to evaluate the PL as a function of the excitation intensity always on the same spot in order to get rid of effects of inhomogeneity. The importance of this aspect can be seen in Figs. 1(a) and 1(b), where the intensity dependent PL is reported for two different position in the sample. Clear differences between the two places are observed.

Looking first at the results in Fig. 1(a), we can see that for the lowest $I_{exc}$ (0.13 mW/cm$^2$) the spectrum is dominated by the LE peak at 1.55 eV, while the HE peak at 1.64 eV is very weak and appears as a shoulder of the former. The LE peak also appears to be very broad, with a wide tail on the low energy side. This tail is probably due to recombination from defect energy levels and we will indicate it as DE. The increase of $I_{exc}$, up to 130 W/cm$^2$, produces the relative enhancement of the HE peak intensity with respect to the LE one. At the same time, the DE band decreases in intensity respect to the other peak and the overall LE emission becomes narrower. In this $I_{exc}$ range both LE and HE peaks blueshift as the excitation increases. When $I_{exc}$ increases further, an inversion of the trend occurs, the LE peak returns to being dominant and both peaks undergo a redshift. We verified that this latter phenomenon is reversible by lowering $I_{exc}$ back down while observing the PL, in order to be sure that we had not damaged the sample, even at the highest excitation densities used here. In Fig. 1(b), where we report spectra from another spot on the sample, the LE and HE peaks are not visible at low $I_{exc}$ and only a single broad peak located at 1.41 eV dominates over the whole spectrum. We note here that at $I_{exc}$=130 mW/cm$^2$, that corresponds closely to the solar constant, the DE recombination is still clearly visible in both spectra (Figs. 1(a) and (b)) taken at low temperature. Of course, a photovoltaic cell works at room temperature, where the DE emission is not visible; but our measurement gives an indication of the relative importance of the defect density at the low excitations typical of sunlight. Increasing $I_{exc}$, this peak blueshifts until it reaches the energy typical of the LE peak, (1.55 eV) at 13 W/cm$^2$. At intermediate I$_{exc}$ (see the spectrum at 1.3 W/cm$^2$) we observe two distinct peaks: LE that emerges at 1.55 eV and DE that shifts, eventually merging with LE. The HE peak needs more excitation density to appear and to gain prevalence over the LE peak with respect to the case of the first spot shown in Fig. 1(a). The overall behavior of the PL taken in the two different spots is hence qualitatively the same but the impact of defects and the power density necessary to better observe the higher energy features is different in the two cases, showing how the inhomogeneity can affect the emission properties of this material.

Finally, it is worth noticing that in the spectra taken at the highest excitation intensities it is possible to identify a low-energy shoulder for both the HE and LE peaks (see black arrows in Fig. 1(a) and (b)). For HE it is possible to see it for I$_{exc}$ between 64 W/cm$^2$ and 190 W/cm$^2$, depending on the illuminated spot. For the LE peak the low-energy shoulder is visible at 380 W/cm$^2$.

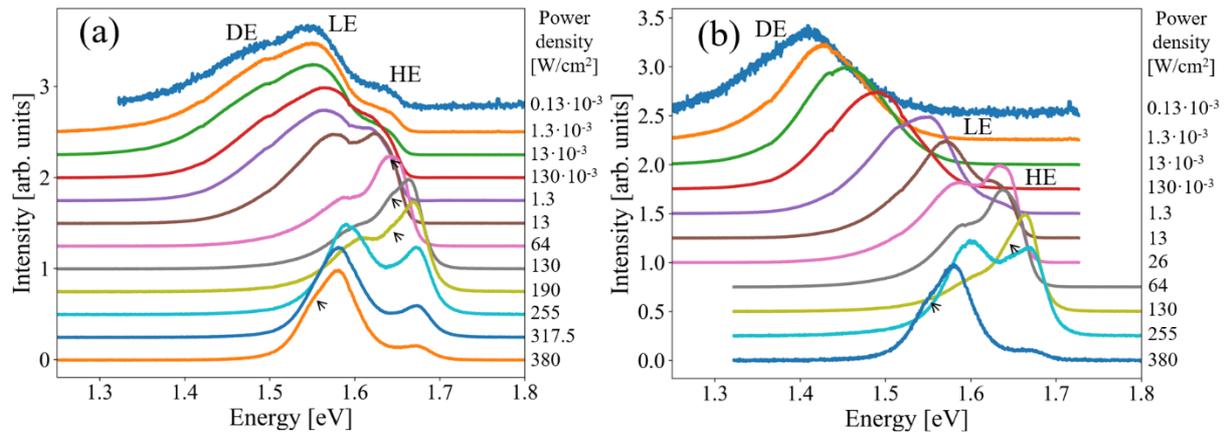

Figure 1: a) PL spectra (normalized to the maximum intensity) of a MAPbI$_3$ thin film as a function of incident excitation power density recorded at 10 K. Arrows indicate the presence of shoulder on the lower energy side of the HE and LE peaks. b) The same measurement for a different spot on the sample, where the DE peak is more visible.

*Temperature dependent photoluminescence with cw excitation.*

PL spectra were recorded every 10 K by the controlled heating of the sample from 10 K until the HE peak disappears, between 120-150 K (depending on excitation intensity). The measurements were performed for different incident $I_{exc}$: 1.3 W/cm$^2$, 13 W/cm$^2$, 130 W/cm$^2$ and 255 W/cm$^2$. The results are shown in Fig. 2 (normalized to the maximum intensity) and in Fig. 3 (non-normalized). For the first two lowest $I_{exc}$, the RT PL is also reported in Fig. 2, but it will be used only as a reference in the discussion and not in the following description of the results. At the highest temperatures, for all the power densities, the LE peak is located at ~1.58 eV, as reported in Fig. 2. By decreasing the temperature, the LE peak blueshifts for $I_{exc}$ =1.3 W/cm$^2$ (Fig. 2(a)), 13 W/cm$^2$ (Fig. 2(b)) and 130 W/cm$^2$ (Fig. 2(c)), while for $I_{exc}$ =255 W/cm$^2$ (Fig. 2(d)) its energy does not change (at most a small redshift is present). As the temperature decreases, the HE peak emerges and gains intensity. The highest temperature at which the HE emission appears, depends on the excitation density: between 1.3 W/cm$^2$ and 13 W/cm$^2$ the HE PL appears at comparable temperatures (140 K) but the temperature lowers for 130 W/cm$^2$ and 255 W/cm$^2$, to 110 K and 100 K respectively. We see that below 100 K the HE peak is favored with respect to the LE peak as $I_{exc}$ increases from 1.3 W/cm$^2$ to 130 W/cm$^2$, while for 255 W/cm$^2$ the LE peak becomes dominant for the whole temperature range and especially at low T. A further power dependent feature, for decreasing T below the value at which the HE peak appears, is the unusual redshift of both peaks that, however, shows a clear dependence on $I_{exc}$. Increasing $I_{exc}$ the shift is considerably reduced for both peaks; for 255 W/cm$^2$, in particular, the HE redshift totally disappears, while for the LE peak even a small inversion to a red shift can be observed. The low-energy broadening of the LE peak, which has been observed at 10 K at low excitation conditions, is present in the spectra taken at $I_{exc}$ = 1.3 W/cm$^2$ (Fig. 2(a)) up to 100 K, while it is difficult to distinguish it for higher $I_{exc}$. Finally, at the highest temperatures, we observe for all excitation intensities the weak shoulder on the low energy side of both LE and HE peaks, observed at 10 K only for the highest powers.

We would also like to highlight a very peculiar characteristic of these temperature dependent PL spectra: the intensity ratio between the HE and LE peaks is not monotonous as a function of T, especially at low $I_{exc}$. To emphasize this aspect, we report colormaps in Fig. 3, where the spectra are not normalized and where it is possible to observe also the intensity trend of the two peaks as a function of the temperature.

From these plots, we can see that the LE peak intensity increases as the temperature is raised up to about 90-130 K, depending on $I_{exc}$, and then decreases. This behavior is smoother at high excitation intensities. In contrast, the behavior of the HE peak intensity is quite different as it exhibits a maximum at decreasing temperatures as the excitation power increases. The integrated intensity of the PL reflects the different behavior of the two main spectral components and shows a temperature dependence strongly related to the excitation conditions. In Fig. 4 the intensity integrated over the whole spectrum is reported as a function of the temperature, for low (1.3 W/cm$^2$) and high (255 W/cm$^2$) power densities. In the first case the intensity rapidly increases with increasing temperature showing a maximum at 130-140 K, while in the second case it slightly increases up to about 80 K and then decreases at higher temperatures.

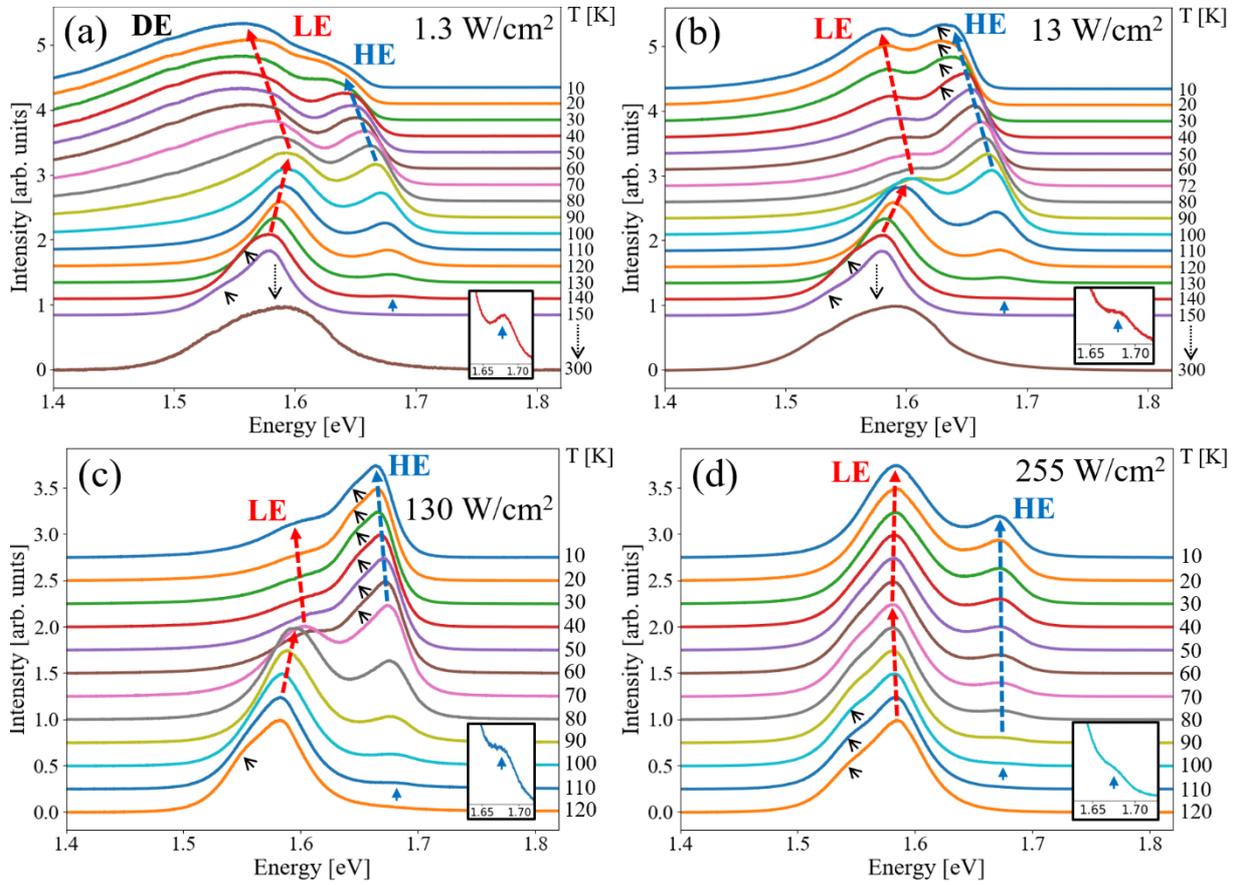

Figure 2: Normalized PL spectra of MAPbI$_3$ thin film as a function of temperature, recorded for different excitation power densities, I$_{exc}$: (a) I$_{exc}$ =1.3 W/cm$^2$, b) I$_{exc}$ =13 W/cm$^2$, c) I$_{exc}$ =130 W/cm$^2$, d) I$_{exc}$ =255 W/cm$^2$). Thin black arrows indicate the presence of shoulder on the low energy side of the HE and LE peaks. Bold blue arrows indicate the temperature at which the HE peak emerges. Red and blue dotted arrows qualitatively indicate the shifts for LE and HE peaks respectively. The inset figure shows a detail of the HE peak first appearance in the spectrum.

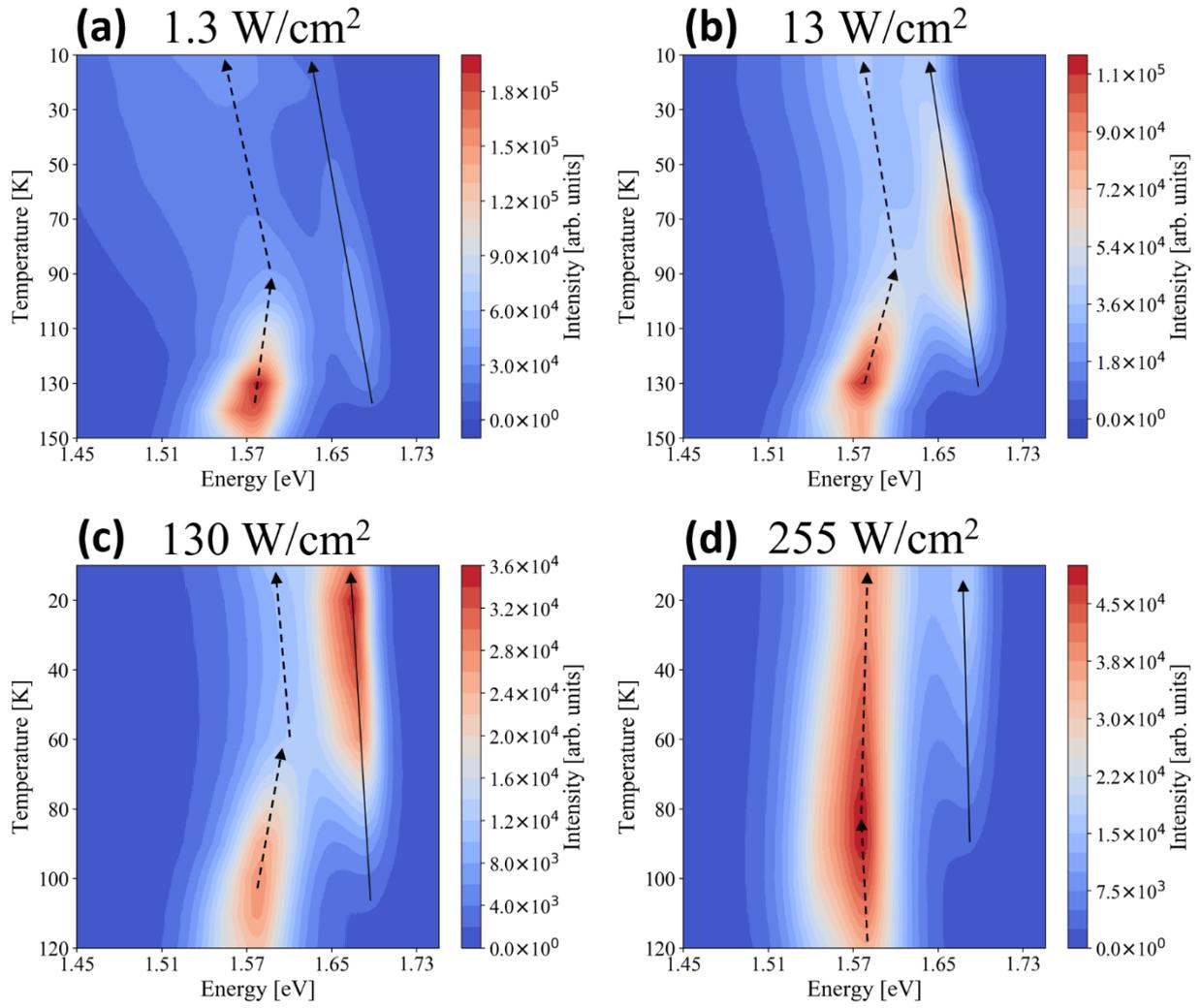

Figure 3: Non-normalized PL spectra of MAPbI$_3$ thin film as a function of temperature, recorded for different excitation power densities (a) $I_{exc}$ =1.3 W/cm$^2$, b) $I_{exc}$ =13 W/cm$^2$, c) $I_{exc}$ =130 W/cm$^2$, d) $I_{exc}$ =255 W/cm$^2$). Dashed and plain lines qualitatively show the trend of the LE and HE peaks, respectively.

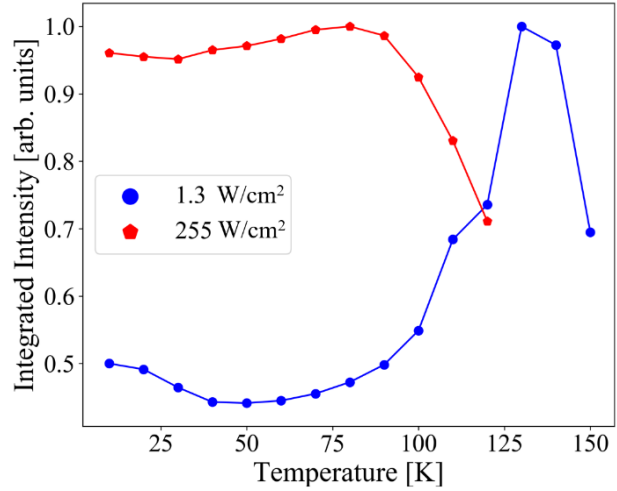

Figure 4: PL intensity integrated over the entire non-normalized spectrum as a function of temperature for different excitation power densities ($I_{exc}$=1.3 W/cm$^2$, blue points, and $I_{exc}$=255 W/cm$^2$, red points; the lines are guides to the eye). The normalization has been performed to show them in the same plot.

*Time-resolved and spectral PL with pulsed excitation.*

To investigate further the recombination processes that produce the different emission features and to compare spectra between two different excitation regimes, time resolved PL (TRPL) and spectral PL measurements were performed under pulsed excitation (Fig. 5).

In Fig. 5(a) we report the PL spectrum recorded with a fluence of 1.13 μJ/cm$^2$ and a temperature of 77 K that shows both HE and LE peaks similar to those measured under cw excitation at 70-80 K. Fig. 5(b) presents the TRPL measurements for both the peaks marked in Fig. 5(a), along with the best fitting model for the data. They have been fit with a double-exponential decay function with two characteristic times (a faster $\tau_1$ and a slower $\tau_2$). This highlights the presence of at least two interplaying recombination mechanisms: usually the faster one is attributed to non-radiative and the slower one to radiative recombination. However, the recombination characteristic times we have found (in particular for $\tau_2$ in the tetragonal phase) are much longer than the usual ones measured for direct bandgap semiconductors, especially considering that a MAPbI$_3$ thin film is particularly subject to the presence of defects. Our result is not unique, long lifetimes are commonly found in the MAPbI$_3$ literature [5]. The decay times are shorter for the HE emission ($\tau_1$ = 1.9 ns, $\tau_2$ = 8.3 ns) than for the LE peak ($\tau_1$ = 3.3 ns, $\tau_2$ = 14.5 ns). The same measurements performed at 5.7 µJ/cm$^2$ (Figs. 5(c) and 5(d)) reveal a faster decay time for both the HE and LE peaks, respectively $\tau_1$ = 0.84 ns, $\tau_2$ = 3.9 ns for HE and $\tau_1$ = 1.85 ns, $\tau_2$ = 10.4 ns for the LE peak. In Fig. 5(c) it is also possible to see a third sharp peak that emerges at ~1.64 eV with a shorter $\tau_1$ decay time (≤0.5 ns, our IRF), while $\tau_2$ remains similar to that of the HE peak (3.6 ns). The last measurement performed at 57 µJ/cm$^2$, shown in Figs. 5(e) and 5(f), reveals the presence of a broad and intense peak at ~1.63eV with a very short decay time, shorter than our IRF.

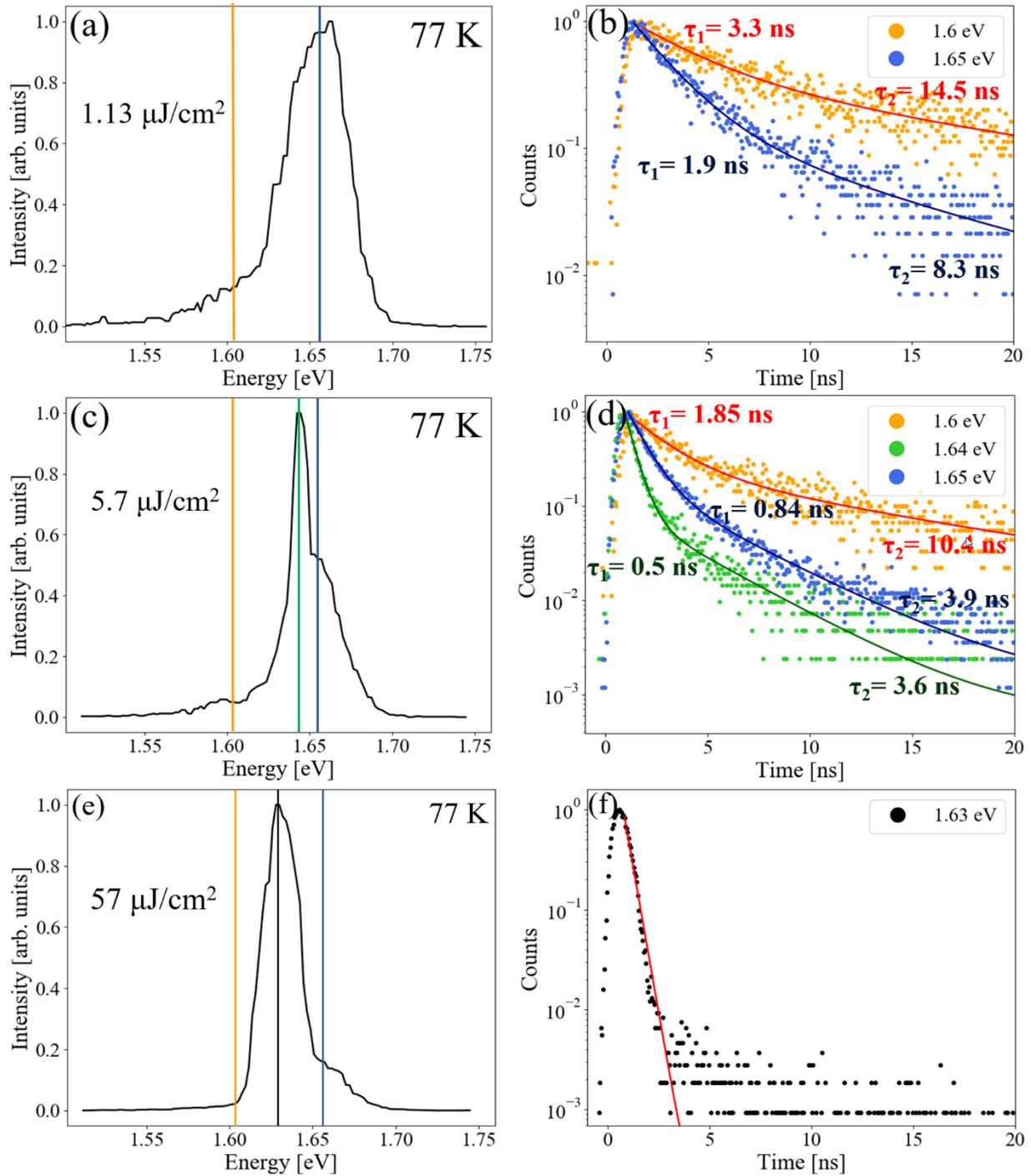

Figure 5: a) Normalized PL spectrum of MAPbI$_3$ thin film at 77 K, excited by a 520 nm pulsed laser (fluence = 1.13 µJ/cm$^2$). LE and HE peak positions are indicated by orange and blue lines, respectively. b) Normalized TCSPC measurement at 1.13 µJ/cm$^2$ for the LE peak (orange) and HE peak (blue) with a double exponential fit (solid lines). c) Normalized PL spectrum with fluence = 5.7 µJ/cm$^2$. LE and HE peaks are indicated by orange and blue lines respectively, while the green line shows the ASE peak. d) Normalized TCSPC measurement at 5.7 µJ/cm$^2$

for the LE peak (orange), the HE peak (blue) and the ASE peak (green) with a double exponential fit (solid lines); the fitted characteristic time $\tau_1$ for 1.64 eV curve is shorter to the IRF (0.5 ns). e) Normalized PL spectrum with fluence = 57 μJ/cm$^2$. f) Normalized TCSPC measurement at 57 μJ/cm$^2$ with single exponential fit; the fitted characteristic time $\tau_0$ is shorter than the IRF (0.5 ns).

## IV. Discussion

Analyzing the peculiar aspects of MAPbI$_3$ PL spectra at low temperature, the most evident characteristic is the double peak emission, visible below the transition temperature from tetragonal to orthorhombic phases. This observation is not new and its principal features have been widely studied, but there is still no agreement in the literature about why it is possible to observe the PL features that seem to belong to both tetragonal and orthorhombic phases at temperatures below the tetragonal to orthorhombic phase transition. Some works [10], [29, 30] propose the presence of inclusions of the tetragonal phase in the orthorhombic one, others [13] attribute the two peaks to bound and free excitons, a further proposal is that the low-energy peak is due to a donor-acceptor pair (DAP) recombination [14] and finally Dar et al. [11] suggest that the two peaks are due to the contemporary presence of ordered and disordered domains of the orthorhombic phase. The presence of tetragonal inclusions within the orthorhombic phase is currently the most common explanation [12, 25], also thanks to the support of independent measurements like XRD. [30] A further feature observed in our measurements is the saturable defect emission (DE) at energies lower than the intrinsic LE peak. These emissions were already observed as an enlargement or shift of LE peak at low excitation [10,12,13] but they were never investigated for a such wide range of $I_{exc}$. We start the discussion of our data from this last point.

When illuminated with low $I_{exc}$ (< 1.3 W/cm$^2$) at T< 70 K (Figs. 1(a) and (b)), the material reveals some emission, that we have indicated with DE, at energies below that of the LE peak.

This DE emission can appear as a tail of the LE band (Fig. 1(a)) or as an independent peak (Fig. 1(b)), depending on the illuminated spot. These DE emissions are likely to be due to the e-h recombination on defect states laying within the bandgap. The number of these defect states is limited, but not negligible (see below), and the DE emission becomes saturated with increasing $I_{exc}$ and hence the density of e-h pairs that recombine at the related energy. We cannot locate these defects precisely in energy because they appear in a wide spectral range, depending on the illuminated spot on the sample and on the light intensity. In any case, they can be located between 1.41 eV and 1.55 eV. PL does not allow the identification of the chemical or structural origin of defect states without a preliminary and systematic study of its properties after the deliberate introduction into the lattice of known doping elements or structural defects [31]. No study of this type on HHPs is present in literature. $MAPbI_3$ perovskite presents indeed many types of possible intrinsic, or light induced, shallow defects.[32].

What we can say from our data is that the blueshift of the DE peak with increasing $I_{exc}$ is compatible with state filling of defect states distributed on an energy band as well as with a donor-acceptor pair emission [33]. The fact that at low $I_{exc}$ the DE signal can be observed up to 100 K (Fig. 2(a)) suggests the former interpretation to be more probable as DAP emissions, which usually disappear at much lower temperatures because their observation is limited by the binding energy of the shallowest of the two involved impurities [33]. The blue shift attributed to band state filling, evident in Fig. 1(b), implies that the defect density is high enough to form an energy level band. Defect levels, indeed, depending on their binding energy and density show different power law dependence, saturate at high enough excitation intensities, possibly leading to one or several crossover points from one power law behavior to another [34]. Low-density defects or impurities would have discrete levels with an ensuing PL energy that would be independent of the power density [29]. This observation gives an important information about defect states in $MAPbI_3$ and their inhomogeneous distribution.

Increasing $I_{exc}$ up to 130 W/cm$^2$ (Fig. 1(a)), and keeping the temperature fixed, the DE emission becomes relatively weaker and an increment of the HE intensity is visible. This feature was already observed [10,11,13,26,28] and generally attributed to the saturation of energetically favored small LE inclusions, similarly to what happens with defects. Indeed, if all the available energy states are filled in the LE tetragonal domains, the carrier transfer from orthorhombic to tetragonal domains is stopped, and more e-h recombination is allowed in the HE orthorhombic domains. An important aspect of our measurements is that the appearance of the HE peak is not solely dependent on the lattice temperature. Its presence is also dependent on the excitation density with a complex interplay with the temperature. The intensity ratio between the HE and LE peaks is generally not monotonic, neither with temperature nor with excitation density. All these features suggest that there is a competitive behavior between the HE and LE electronic states. This can be due to a fast carrier transfer from HE to LE domains as the latter are energetically favored [10, 11, 14, 29, 30]. In this way, even if the LE inclusions are not widespread, they strongly contribute to the PL, reducing the recombination in HE domains. This conclusion is also supported by TCSPC measurements that show a shorter lifetime for HE emissions than for the LE ones.

Let us now discuss the dependence of the LE and HE energies upon temperature (Fig. 2). The often observed redshift of the two peaks for decreasing temperature below 80-100 K, which is an uncommon feature in high quality semiconductors but it can be observed in defected materials [35], is explained with reverse ordering of the band edge.[15] However, if we carefully look at the data reported in Fig. 2, we notice that this red shift is strongly reduced and eventually vanishes as $I_{exc}$ is increased (Fig. 2(c) and (d)), approaching an ordinary semiconductor behavior. At the highest $I_{exc}$ (Fig. 2(d)) we even observe a small blue shift for the LE peak below 70 K. High excitation intensities give rise to an intrinsic regime of the material, in which the contribution to the PL of defects and impurity states is negligible. If we look at the temperature dependence of the PL intensity (Fig. 4) we observe the usual behavior of semiconductors again for high excitations: in this case the integrated intensity only slightly increases up to a certain T (70-80 K for MAPbI$_3$) and then decreases at higher T. On the other hand, for lower I$_{exc}$, the PL intensity has a strong non-monotonic dependence on T that is anomalous for ordinary semiconductors. These features suggest that the presence of defect or impurity states and also of weakly localized states due to compositional disorder [36] are favoured at low T and are detected at low I$_{exc}$ thus shifting the PL peaks towards lower energies. The importance of localized states in this excitation range is also supported by the increase of the PL intensity for increasing temperature, as intrinsic, extended e-h states are less affected by non-radiative recombination. The fact that the HE peak intensity appears more affected than that of the LE peak is then attributed to the fact that LE states act as trapping states for e-h pairs excited in the orthogonal domains. However, it is also important to highlight that the latter observations may be affected by a local heating of the sample, induced directly by the excitation source at high I$_{exc}$, an aspect that we will discuss later. Of course, we are aware that the further increase of the temperature eventually leads to a red shift. This is also seen in Figs 2(a) and 2(b) where we have reported the RT PL for comparison. As a matter of fact, however, the RT PL is very broad, much broader than at 150 K. This broadening cannot be explained by simple thermal factors, because it is much larger than the difference in kT. Different spectral features contribute to the RT spectra, whose origin is not self-evident and will be not discussed here.

All the features observed as a function of temperature and excitation intensity strongly suggest that the LE peak is due to radiative recombination in tetragonal inclusions in the orthorhombic lattice. Indeed, the LE features vary very smoothly with T and I$_{exc}$ strongly suggesting that there is no change in its nature during the variation of the lattice temperature.

Above 130 W/cm$^2$, the LE peak starts growing again and returns to be dominant. This is an unexpected result and seems to tell us that more LE domains are induced by higher I$_{exc}$. This trend is similar to the evolution in temperature for low power densities and suggests that the intense incident light can locally heat the sample, thermally promoting the tetragonal phase and relative emission. Further features, observed in the temperature dependent measurements, support this hypothesis: we can indeed see that for 130 W/cm$^2$ and 255 W/cm$^2$ the HE peak is only observed at the lowest temperatures, in contrast to what we should observe according to the picture described above regarding the interplay between the LE and HE intensities. The observation suggests that the light locally heats the sample and that the real temperature under the laser spot is higher than the surrounding lattice temperature. Panzer et al. [20] reported similar evidence of low temperature promotion of the tetragonal PL peak, induced by high fluence pulsed excitation, and explained it by evoking local heating. In their case, however, after high fluence pumping, the peak persists even if the same spot is probed with lower fluences, while in our case the system completely recovers its previous emission characteristics on lowering the power. The peak they observe, moreover, is narrower (10-15 meV) and they propose amplified spontaneous emission (ASE) as its origin. This different behavior reversing the excitation density can be due to the two distinct regimes of excitation used (continuous in our case, pulsed in theirs): for the same amount of average power, a pulsed light will transfer all the energy during the time interval of the pulse, but the local heating is more difficult because of the inactive part of the duty cycle. For these reasons, with pulsed light, the higher energy density promotes a higher e-h pair density with a higher possibility to induce population inversion during the pulse, activating ASE. As a matter of fact, Panzer and coworkers [20] report ASE for the tetragonal phase also at room temperature, and for the orthorhombic phase at low temperatures. We do not detect these kinds of sharp peaks in our cw measurements, but we have observed low energy shoulders in the spectra recorded at high excitations for both LE and HE peaks. Their energy position (low energy side) and separations (~ 20-30 meV) with respect to the main peaks, moreover, are comparable with those observed in ref. [20] and with an ASE. On the other hand, we observe a very sharp peak in TCSPC measurements at 5.7 μJ/cm$^2$ and 57 μJ/cm$^2$. Its energy position is compatible with the shoulders observed in cw regime and its decay time is shorter than the other observed emissions. We cannot estimate decay time because it is shorter than our IRF (0.5 ns) but this feature is compatible with an ASE. The similarity between the emissions observed in the pulsed regime with the low energy shoulders that appear on the low-energy side of the HE and LE peaks suggests that gain conditions can be reached in these samples also under cw excitation, even if a local thermal heating under the light spot occurs.

## V. CONCLUSION

In conclusion, we performed a systematic study of the PL emission of MAPbI$_3$ perovskite thin films at low temperatures, using cw excitation intensities varying over about seven orders of magnitude with the use of extremely low excitation intensities, obtaining a coherent set of data, to better understand the complex and elusive emission characteristics of this material. We have demonstrated the significant influence of incident light intensity on the qualitative and quantitative response of the MAPbI$_3$ thin film. In particular, we have shown that at low T (< 100K) and low $I_{exc}$ (< 1.3 W/cm$^2$) the presence of defect-related emission can be easily observed, also highlighting the relevant differences that can be observed within a single sample due to its inhomogeneities. The power and temperature dependence of the defect band suggests that even in high quality MAPbI$_3$, capable of providing high photovoltaic efficiency, the defect density is high enough to give rise to an energy level band. We have also shown that the intensity ratio between the two components of the doublet usually observed below 160 K, LE and HE, is a complex function of T and $I_{exc}$ as carrier collection in the low energy regions and its saturation determine that ratio. The same analysis has allowed us to observe that under excitation conditions that favor intrinsic recombination, the behavior of MAPbI$_3$ approaches that of ordinary inorganic semiconductors. Indeed, the peaks redshifts are reduced as well as the non-monotonicity of PL integrated intensity. Finally, our systematic measurements strongly suggest that the LE peak is due to recombination in tetragonal inclusions into the orthorhombic lattice where the recombination that gives rise to the HE emission occurs.

At high excitations ($I_{exc}$ > 13 W/cm$^2$), we have shown that ASE is attainable also in cw conditions, for which a local, non-destructive, heating of the lattice occurs. This result should warn against the use of high $I_{exc}$ excitation to characterize this material and demonstrates that characterization studies should be performed with cw excitation at low power densities and low temperatures.

The last result of our work is that the widely contrasting results shown in the literature can be explained, apart from the intrinsic inhomogeneous nature of MAPbI$_3$ samples, by the fact that temperature and excitation power vary from paper to paper and that the use of pulsed excitation can be misleading. This is in particularly true with regard to the features of the PL doublet usually observed at low temperatures, for which it is unrealistic to draw quantitative conclusions using only one excitation condition.

## ACKNOWLEDGMENTS

This work has received funding from the Horizon 2020 program of the European Union for research and innovation, under grant agreement no. 722176 (INDEED).

## REFERENCES

*valerio.campanari@uniroma2.it

*faustino.martelli@cnr.it